\documentclass[aps,prl,twocolumn,showpacs]{revtex4-1}

\usepackage{graphicx}
\usepackage{dcolumn}
\usepackage{bm}
\usepackage{epsfig}
\usepackage {epstopdf}
\usepackage{subfigure}

\begin{document}

\title{Retention capacity of random surfaces}

\author{Craig L. Knecht$^1$}
 \email{craigknecht03@gmail.com}
\author{Walter Trump$^2$}
\email{w@trump.de}
\author{Daniel ben-Avraham$^3$}
\email{qd00@clarkson.edu}
\author{Robert M. Ziff$^4$ }
\email{rziff@umich.edu}
\affiliation{$^1$ 691 Harris Lane, Gallatin TN  37066 USA}
 \affiliation{$^2$ Department of Physics, Gymnasium Stein, 90547 Stein, Germany}
 \affiliation{$^3$ Department of Physics, Clarkson University, Potsdam NY 13699-5820 USA}
\affiliation{$^4$ Michigan Center for Theoretical Physics and Department of Chemical Engineering, University of Michigan, Ann
Arbor MI 48109-2136 USA}

\date{\today}

\begin{abstract}

We introduce a ``water retention" model for liquids captured on a random surface with open boundaries,
and investigate the model for both 
continuous and discrete surface heights $0, 1, \ldots n-1$
on  a square lattice with a square boundary.
The model is found to have several intriguing features, including a nonmonotonic 
dependence of the retention on the number of levels: for many $n$,
the retention is counterintuitively greater than that of an $n+1$-level system.  The behavior
is explained using percolation theory, by mapping it to a 2-level system with variable probability.  
Results in one dimension are also found.
\end{abstract}
\pacs{64.60.ah, 64.60.De, 05.50.+q}

\maketitle

\begin{figure}[bthp]
\centering
\includegraphics[width=3.5 in]{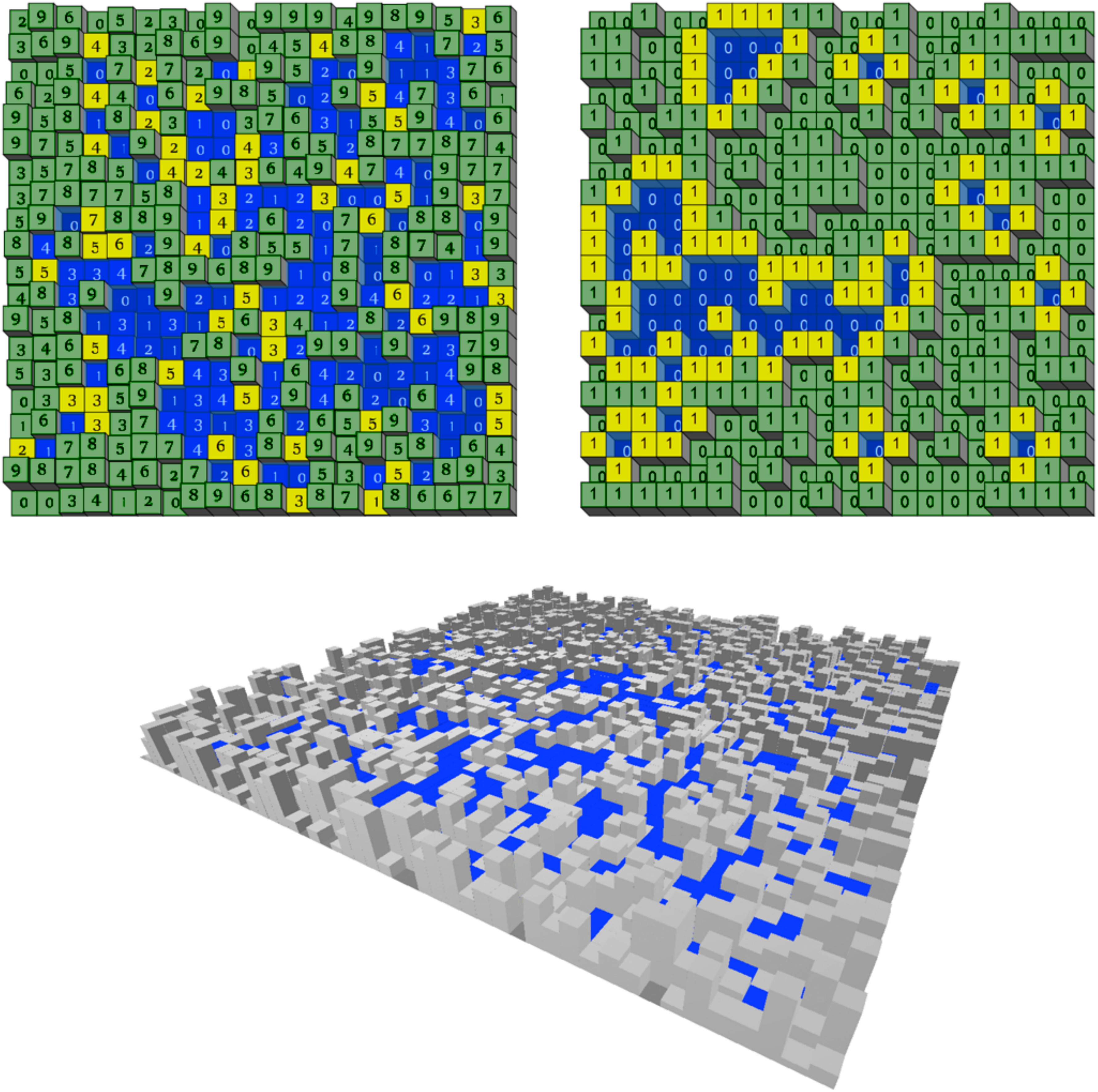}
\caption{(color online) Examples of water retention on square systems with an equal distribution of terrain heights.
Top left: heights 0 through 9; Top right: heights 0 and 1.  Green (gray with black height numbers)
are dry and above water,
blue (gray with withe numbers) are filled with water, and yellow (light shade) are spillover sites of the height of 
an adjacent pond (and only shown if neighboring a wet site).  Bottom: Perspective view of a water-filled 10-level system,
with dry sites shown as grey (lighter shading on top).
}
\label{twolevel}
\end{figure}

\vskip 1 in
Consider a bounded horizontal random surface with a landscape of varying height,
as shown in Fig.\ \ref{twolevel}.
A liquid such as water is dripped over the surface and is allowed to drain out all of the 
boundaries.  Internal sites in valleys capture the
water and create ponds, and eventually all the ponds fill up to their maximum height.  
We are interested in finding the total amount of water retained 
in the system when the maximum heights are reached.  Physically, this problem is related to coatings
on a random surface and the properties of landscapes and watersheds.
Theoretically, it is related to the topology of random surfaces~\cite{MajumdarMartin06,CarmiKrapivskybenAvraham08} and to invasion percolation (IP), but with some interesting new features.

We study this problem on a regular square lattice with random heights assigned
to each site.  The systems are square of
size $L \times L$ with draining boundaries on all four sides.  Extensive simulations were
performed with uniformly distributed discrete heights  $0, 1, 2, \ldots n-1$ for values of $n$ ranging from 2 to 100,
and also for a continuum of heights $0, 1$.  We also studied a 2-level system with variable occupation
probabilities of the 2 heights. The simulation method we used
is a form of  IP  in which we  effectively reversed the flow and flooded
the system from the outside with higher water levels, 
and recorded the level of the water in a pond when it was first flooded.
The retention is the difference between that level and the height of
the terrain below the pond.  

\begin{figure}[htbp]
\centering
\includegraphics[width=3in]{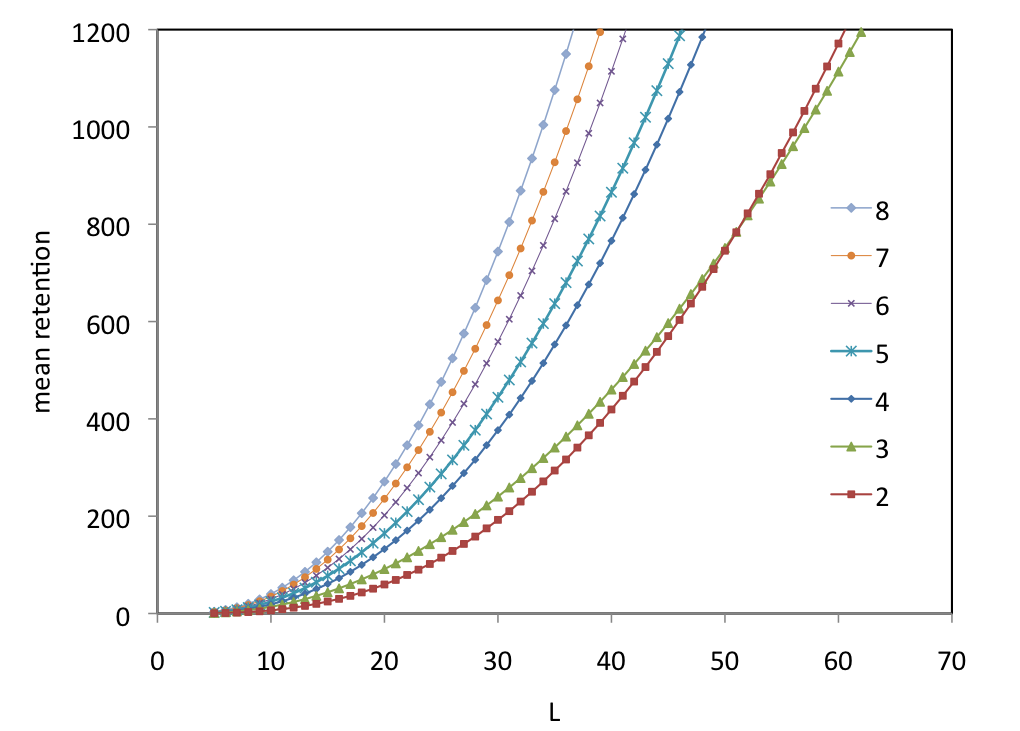}
\caption{(color online) Retention $R_n^{(L)}$ vs.\ $L$ for $n = 2, 3, \ldots, 8$ 
with a uniform distribution of levels, showing the
crossing of the curves for $n = 2$ and $3$ at $L \approx 51$. 
Additional crossings are found for larger $L$ (see Table \ref{crossingtable}).
For small $L$ there is an ordering of the retention with
$n$, which is violated for larger $L$.}
\label{retentiontwotoeight}
\end{figure}

Fig.\ \ref{retentiontwotoeight} shows the average retention
$R_n^{(L)}$ on $n$-level systems, for $n = 2, \ldots 8$,
as a function of $L$.
Here we assume that all terrain heights
occur with equal probability.
As expected, the retention grows as $L^2$ for large $L$,
and generally grows with $n$, as more levels create
deeper ponds.   However, we found
deviations to this expected behavior.  As  seen in 
Fig.\ \ref{retentiontwotoeight},  there is a crossover in the curves
for $n = 2$ and $3$: for small $L$,
$R_2^{(L)} < R_3^{(L)}$, but for $L > 51$, 
$R_2^{(L)} > R_3^{(L)}$.
This is in spite of the fact that 3-level systems can have ponds of water
of depth 2, while 2-level systems 
can only have ponds of depth 1.  
Further study shows additional crossings between levels $n$ and $n+1$ at
seemingly random $n$'s, and at  larger values of $L$ (Table \ref{crossingtable}).

In this Letter we explain some of the puzzling features of this model, though many questions remain.
Some related issues, especially involving multi-level nonuniform systems, are discussed in \cite{BaekKim11}.

To analyze the multi-level discrete model, we  make a  decomposition of $R_n^{(L)}$ in terms of the  retention in a 2-level system with varying $p$,  $R_2^{(L)} (p)$,
where $p =$ Prob$(0)$ is the probability or fraction of sites with terrain height 0 in the 2-level system:
\begin{equation}
R_n^{(L)} = \sum_{i = 1}^{n-1} R_2^{(L)}\left({ \textstyle \frac{i}{n}}\right)\,.
\label{decomposition}
\end{equation}
The $i = 1$ term represents the amount of water retained
up to just the first level, for which all sites of terrain height 1 or higher can be considered as level 1.  The net fraction of 0-height 
sites is  $1/n$.  The $i = 2$ term represents the total
amount of water at just the second level; here we collapse the first
2 levels into the new level 0 (fraction $2/n$), and the rest of the
sites can be considered as level 1.  Likewise, the remaining terms
follow.

\begin{figure}[htbp]
\centering
\includegraphics[width=3in]{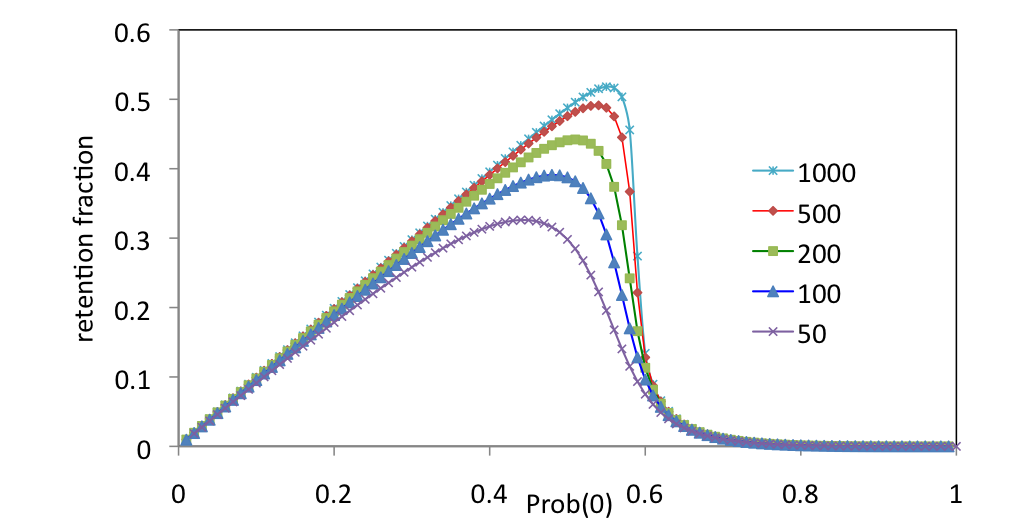}
\caption{(color online) Retention fraction $R_2^{(L)}(p) / L^2$ in a 2-level system with varying $p=$Prob$(0)$, for $L$ ranging from 50 to 1000, averaged
over 10\,000 to 1\,000\,000 samples for each point.}
\label{twolevelretention}
\end{figure}

It is also possible to show that $(1/i)R_2^{(L)}(i/n)$ is equal to the number of sites with retention $i$.
Thus, the total number of wet sites is
\begin{equation}
W_n^{(L)}  = \sum_{i=1}^{n-1} \frac{1}{i}R_2^{(L)}({\textstyle \frac{i}{n}}) \,.
\label{wetsite}
\end{equation} 
%
%
It is therefore necessary to just know the behavior of the 2-level system with a varying $p$ in order to predict the behavior of  all $n$-level systems
(including those with nonuniform level distributions).
We have carried out simulation of $R_2^{(L)}(p)$ for $p = 0.01$, 
0.02, $\ldots, 0.99$ for various $L$, and the results are shown in Fig.\ \ref{twolevelretention}. For small $L$,
the curve is rounded and peaked close to $P(0) = 1/2$, but for larger  $L$ it approaches a ramp of slope 1,
up to the value of $p = p_c = 0.592\, 746$ (the site percolation threshold), after which it drops off 
precipitously and rapidly approaches 0.    This behavior is best understood from the limit $L\to\infty$.
Define $r_2(p) = \lim_{L \to \infty} R_2^{(L)}(p)/L^2$. In an infinite system, all finite clusters of $0$-sites
retain water and the infinite cluster alone drains off.  Thus, the total retention is
\begin{equation}
r_2(p) = p- P_{\infty}(p),
\label{r2}
\end{equation}
where $P_{\infty}(p)$ is the fraction of sites belonging to the infinite cluster.
Very close to (and above) $p_c$, $P_\infty \sim  a(p-p_c)^\beta + \dots$ where $a$
is a constant and $\beta = 5/36$ \cite{StaufferAharony94}.
Because $P_\infty$ rises quickly as $p$ increases beyond $p_c$, we get the quick drop to zero in $r_2(p)$.
For finite $L$, finite clusters at the boundaries drain as well, yielding the observed finite-size effects.

Exactly at $p_c$, the drainage area is fractal, yet the retained water
is still proportional to $L^2$ for large $L$ with corrections proportional to $L^{d_f}$ where
$d_f = 91/48$ is the fractal dimension.  We verified that at $p_c$, the size distribution
of the draining clusters (boundary clusters in percolation) satisfies  $n'_s \sim s^{-\tau'}$
with $\tau' = 1/d_f - 1 = 139/91 \approx 1.527$ as predicted in \cite{LookmanDeBell92}.
Our measurement of $\tau' = 1.5256 \pm 0.003$
confirms this prediction to about 20 times the precision given in \cite{LookmanDeBell92}.

As a good approximation for large $L$, we can ignore
the small contribution to $R_2^{(L)}(p)$ for $p>p_c$ and
approximate $R_2^{(L)}(p) = p L^2$ for $p < p_c$.  Then, from
(1), we find the following formula for the $n$-level retention
in the large-$L$ limit:
\begin{equation}
R_n^{(L)}/L^2  \approx \sum_{i=1}^{n^*} \frac{i}{n} = \frac{n^*(n^*+1)}{2n}
\label{daniequation}
\end{equation}
where $n^*$ is the largest integer such that $n^*/n$ is less than
$p_c$.  Thus, we have $R_2^{(L)} \sim (1/2) L^2$, and 
$R_3^{(L)} \sim (1/3) L^2$, which indeed gives 
$R_2^{(L)} > R_3^{(L)}$ for large $L$.  This result can be explained simply
by the fact that for the 2-level system, roughly half
the sites are 0's and filled with water, while for the 3-level system, only
1/3 of the sites are 0's; very
few ponds are filled to a level of 2 because
those sites correspond to clusters above the percolation 
threshold.

To explain the crossing, we must also explain why
the curves for $R_n^{(L)}$ are ordered 
$R_2^{(L)} < R_3^{(L)} < R_4^{(L)} ... $ for small $L$.
This can be understood qualitatively from the
behavior of $R_2^{(L)}(p)$ for small $L$ as
in Fig.\ \ref{twolevelretention}:
because those curves are smooth, equation (\ref{decomposition})
will be a gradual, increasing function of $n$.  To be more
rigorous, we consider the smallest system possible:
a $3 \times 3$ system, which has only one site that
can hold water (the center site), and only four sites 
that can block it, as the corner sites
are irrelevant.  
A direct calculation yields
\begin{equation}
R_n^{(3)} = ( n ^2 - 1)(3  n ^2 - 2)/(60  n ^3) \,,
\end{equation}
which is a monotonically increasing function of $n$.
(Details of the derivation will be given in a future paper.)
Because the ordering is verified for $L = 3$ but 
not  for large $L$,
crossing must necessarily occur for some $L$.

The curves that are ``out of order" and cross
are those in which $r_n = n^*(n^*+1)/(2n)$ is greater than
$r_{n+1}$, by (\ref{daniequation}).  This occurs when the fractional part of $p_c n$ is
between 0 and $1-p_c \approx 0.407$.
 The crossing curves $(n, n+1)$ are at
$(2,3)$, $(4,5)$, $(7,8)$, $(9,10)$,  $(12,13)$, $(14,15)$, $(17,18)$, etc.
We have verified the first six  crossings as shown
in Table \ref{crossingtable}.  For $n > 30$, the
simple analysis based upon (\ref{daniequation}) evidently breaks down as contributions
from $R_2(p)$ for $p > p_c$ become important, and the
crossings are predicted to become less frequent, though we have not
measured them directly.
 
\begin{table}[htdp]
\caption{Crossing points where $R_n^{(L^*)}=R_{n+1}^{(L^*)}$, 
extrapolated to non-integer $L^*$.}
\begin{center}
\begin{tabular}{|l|l|l|}
\hline
$n$ and $(n+1)$	&	$L^*$	&	$R_n^{(L^*)}$	\cr
\hline
2 and 3	&	$51.12$	&	790 	\cr
4 and 5	&	$198.1$	&	26\,000 	\cr
7 and 8	&	$440.3$	&	246\,300 	\cr
9 and 10  &  559.1 & 502\,000   \cr
12 and 13  &  1390.6 & 4\,288\,500   \cr
14 and 15  &  1016.3 &  2\,607\,000   \cr
\hline 
\end{tabular}
\end{center}
\label{crossingtable}
\end{table}%

In the limit that the number of 
levels becomes infinite, the discrete system goes over to the
continuum one.  Now, as in traditional
 IP \cite{WilkinsonWillemsen83}, the fluid flows over the lowest barrier site on the 
perimeter of a pond.   For a continuum bond IP system, the ``raining" IP problem 
has recently been considered  in  \cite{vandenBergJaraiVagvolgyi07,DamronSapozhnikovVagvolgyi09}, 
and the pond-size distribution, away from the boundaries, was investigated.

\begin{figure}[htbp]
\centering
\includegraphics[width=3in]{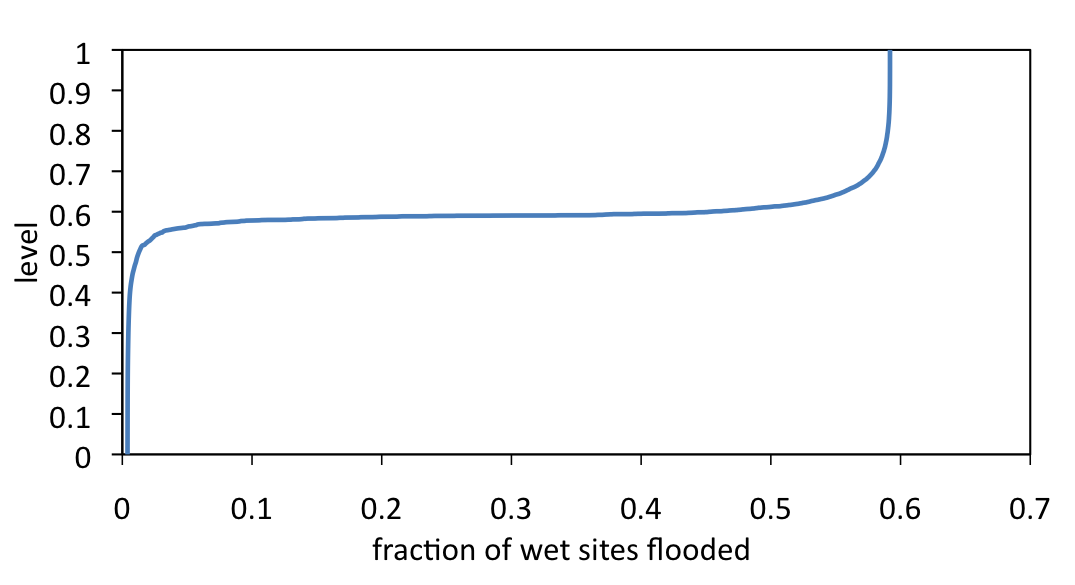}
\caption{Level when sites are first flooded vs.\ the fraction of sites flooded on a $1000 \times 1000$ continuum system.}
\label{levelvsflooded}
\end{figure}

Here, considering the continuum site system, we find that
water rises to an average height of $\overline h \approx 0.6039$ (averaged over wet sites only,  for $L \to \infty$), which is slightly
above $p_c$.  The large ponds have a water level
that is slightly below $p_c$, because higher levels produce
large percolation clusters that would run into the boundary.
There are also small ponds with higher levels, corresponding
to clusters above the threshold; these allow the average
water level to be above $p_c$.   Fig.\ \ref{levelvsflooded} shows the
water level of sites when first flooded as a function of the number of sites flooded,
showing the small contribution of the ponds of high level.

In fact, taking the continuum limit of (\ref{decomposition}) we can calculate the total retention  $r$ per site in the continuum system directly
by integrating the curve of $r_2{(p)}$, $r =  \int_0^1 r_2(p) dp$.
The triangular part below $p_c$ gives $p_c^2/2$ exactly, and
the tail above $p_c$ gives a small correction. The tail's
area extrapolates to $0.0063$ for large $L$, and 
predicts $r = p_c^2/2 + 0.0063 = 0.1820 $, which we verified directly to $\pm 0.0002$                                                                                                                                                                                                                                                                                                                                                                                                                                                                                                                                                                                                                                                                                                                                                                                                                                                                                                                                                                                                                                                                                                                                                                                                                                                                                             by measuring the retention  for systems of up to $L=12\,000$ and extrapolating
to $L = \infty$.  The retention per site $r$ is equal to $\langle w h \rangle / 2 \approx w \overline h / 2 \approx \overline h^2/2$ where 
$w=W^{(L)}/L^2$. 
Note, $w  = \int_0^1 (r_2(p)/p) dp$
follows from (\ref{wetsite}) in the continuum limit, and we find
$w = p_c + 0.0100 = 0.6028$ in agreement with direct measurement (see (\ref{asymptotic}));
 $\overline h = 2r/w = 0.6039$ was also independently measured.

In fact, applying (\ref{r2}) to Eqs.\ (\ref{wetsite}) and (\ref{decomposition}) we see that 
$w=1-\mu_{-1} $, $r=1/2-\mu_0$, and more generally, we find the moments of the retention as
\begin{equation}
\langle r^q \rangle =\lim_{L\to\infty}\frac{\langle(R^{(L)})^q\rangle}{L^2}=\frac{1}{q+1}-\mu_{q-1}\,,
\end{equation}
where $\mu_q=\int_0^1 x^q P_{\infty}(x)\,dx$ is the $q$-th moment of $P_{\infty}$.
Thus, we have found that the moments of $P_{\infty}$ assume a specific physical interpretation
in the context of the retention problem.

The asymptotic behavior of $W^{(L)}$ and $R^{(L)}$ in the continuum system is found to be
\begin{eqnarray}
W^{(L)} &\sim& 0.6028 L^2 - 2.4L^{1.25} \cr
R^{(L)} &\sim& 0.1820 L^2 - 1.27L^{1.25}
\label{asymptotic}
\end{eqnarray}
where the second terms reflect the effects of the
drainage sites and ponds of lower water level near the boundary.
At each level in a discrete system, the drainage area is just all the
clusters touching the boundary, which extends into the system a distance
of the correlation length $\xi \sim |p - p_c|^{-\nu}$. Integrating this over $p$
up to $p_c - c L^{-1/\nu}$ and multiplying by the perimeter $4L$ 
gives a depletion zone $\propto L^{2-1/\nu} = L^{5/4}$.
The value of the  exponent 1.25 was verified numerically to $\pm 0.05$.
Recently, it has been shown that watersheds,
bounded by the ``continental divide" between drainage regions, have a similar
fractal dimension $d_f \approx 1.22$ \cite{FehrKadauArajjoAndradeHerrmann11}.
It appears that these two problems, however, are different, despite the similarity of
exponents.

The assignment of a terrain height for each site using a probability corresponds to
a grand canonical type of description.  One can also distribute
the levels canonically, with exactly $1/n$ of them of each height. We  carried
out simulations using this
ensemble and found only small differences.  
 For $L = 3, 4$, and 5,
we also carried out an exact enumeration of all canonical states.
For small $L$, $\Delta = R_n^{(L)}[\mathrm{canonical}] -
R_n^{(L)}[\mathrm{grand canonical}] > 0$, and for larger
$L$, $\Delta$ decreases.
For $n = 2$, $\Delta$ appears to approach 0,
while for $n = 3, 4, 5$, $\Delta$ appears to approach
a negative constant for large $L$. Because the value of the
retention itself grows as $L^2$, the relative difference $\Delta/R_n^{(L)}$
is very small.  We verified that changing to 
the canonical ensemble does not affect the crossing
behavior of the $R_n^{(L)}$ curves.

 We studied the distribution $n_s$ of ponds of $s$ sites
and verified that the system self-organizes to the percolation critical
point with $n_s \sim s^{-\tau}$ and $\tau = 187/91$.
Unlike standard percolation, we cannot
write exact formulas for any $n_s$---not even for $n_1$.  
However, we can make an estimate for $n_1$ as follows:
The probability that a site is in a pond of size 1,
of water height between $x$ and $x + dx$, is given approximately by
\begin{equation}
P_1(x) dx =4 (1-x)^3 x [1- (1-x)^3]  dx
\label{Ph}
\end{equation}
where the factor of 
$4(1-x)^3$ is the probability that 3 of the neighbors are of higher terrain height
(4 possibilities), 
$x$ is the probability that the site itself is of terrain height less than or equal to $x$,        
and $[1 - (1-x)^3]$ is the probability that the spillway site has at least
one neighbor lower than $x$, so the spillway can drain at least to the next sites.
This gives 
$n_1 \approx \int_{x_0}^1 P_1(x) dx = 0.01624$
where $x_0 = \overline h = 0.6039$ is the average water height surrounding
the cluster.  This compares to a measurement of $n_1 = 0.015\,95$.
Likewise,  the average water height of the ponds of size 1, 
$\overline h_1 = \int_{x_0}^1 x P_1(x) dx/\int_{x_0}^1 P_1(x) dx= 0.6904,$
is  close to the extrapolated measured value 0.6887.

We also studied the model in one dimension (1D), where there are no crossings,
however exact results for all quantities can be found.
Consider, for example, the semi-infinite line (sites $1,2,3,\dots$) so that water can spill only
through the left edge, and assume a uniform distribution of barriers in $[0,1]$.
As we look at the 1D ponds, starting from the left edge, the water level keeps rising the farther we venture
into the line.  In fact, each pond begins when a record-height barrier is encountered, and ends when the next, yet higher
barrier, is met.

The probability that the barrier at site $k+1$ is taller than all the $k$ preceding barriers is $\int_0^1x^k\,dx=1/(k+1)$.
This is also the probability that a pond starts (or ends) at site $k+1$.  Because the barriers demarcating the ponds
occur with probability $1/k$ at site $k$, it follows that the typical size of ponds, $k$ sites away from the edge, is $k$.
The ponds grow linearly with their distance from the edge.

Next consider $p_s(k)$, the probability that a 1D pond of size $s$ is $k$ sites away from the edge, in sites $k+1,k+2,\dots,k+s$.
For that pond to have water level $x$, the first $k-1$ sites must have barriers lower than $x$ (with probability $x^{k-1}$),
as do sites $k+1,k+2,\dots,k+s$ (probability $x^s$).  Site $k$ contains a barrier of height $x$ (probability $dx$), and
site $k+s+1$ contains a barrier of height $y>x$ (probability $1-x$).  Thus the probability  for a pond of level $x$ 
is $x^{k-1+s}(1-x)\,dx$.  Integrating over $x$, we obtain the required probability:
\begin{equation}
p_s(k)=\int_0^1x^{k-1+s}(1-x)\,dx=\frac{1}{(s+k)(s+k+1)}\,.
\end{equation}
Note that $\sum_{s+1}^{\infty}p_s(k)=1/(k+1)$, consistent with our previous result, and that the moments of $p_s(k)$
diverge, which is why we estimated the typical pond size instead.

Similarly, the probability for having $k$ draining sites at the edge is 
\begin{equation}
p_\mathrm{drain}(k)=\frac{1}{(k-1)!}\int_0^1x^k\,dx=\frac{k}{(k+1)!}\,.
\end{equation}
These results illuminate the analogous quantities in 2D, where however no exact results could be found.

In this Letter we have only touched upon the questions that one may ask about the retention model.  There are many more questions
that are unsolved, including the exact results for the size distribution of the clusters, the average retention as a function of the
distance from an edge, the behavior on other lattices, on systems with different boundary shapes, in higher dimensions, and systems with a tilt.
We believe it is an interesting model that warrants much further study. 

We mention finally that the water retention problem was previously studied in the context of surfaces created by magic squares
\cite{Knecht}.  The application to 
random surfaces is an example of the deeper connections of this problem.

{\it Acknowledgments:} The authors  acknowledge correspondence with Neal Madras, Gareth McCaughan, and Seung Ki Baek.
Assistance from Spencer Snow and Joe Scherping is also noted.

\begin{widetext}
Note added:  This paper was published in Physical Review Letters {\bf 108}, 045703 (January 25, 2012).  
Some additional material is given in the web page

\url{http://en.wikipedia.org/wiki/Water_retention_on_mathematical_surfaces},

\noindent where retention on magic squares is also discussed.  Measurements of the crossing points $R_n^{(L^*)}=R_{n+1}^{(L^*)}$ for several larger values of $n$  have been found and will be given in a future publication.  

\end{widetext}


\begin{thebibliography}{99}

\bibitem{MajumdarMartin06} 
S. N. Majumdar and O. C. Martin,
Statistics of the number of minima in a random energy landscape,
Phys. Rev. E {\bf 74}, 061112 (2006).


\bibitem{CarmiKrapivskybenAvraham08}
S. Carmi, P. L. Krapivsky, and D. ben-Avraham,
Partition of networks into basins of attraction,
Phys. Rev. E {\bf 78}, 066111 (2008).



\bibitem{BaekKim11} S. K. Baek and B. J. Kim, arXiv 1111.0425.

\bibitem{StaufferAharony94} D. Stauffer and A. Aharony, Introduction to Percolation Theory (Taylor and Francis, Philadelphia, 1994).

\bibitem{LookmanDeBell92} T. Lookman and K. De'Bell, Surface fractal dimension for percolation clusters: A Monte Carlo study, Phys. Rev. B {\bf 46}, 5721 (1992).

\bibitem{WilkinsonWillemsen83} D. Wilkinson and J. F. Willemsen, Invasion percolation: a new form of percolation theory. J. Phys. A  {\bf 16} 336 (1983).

\bibitem{vandenBergJaraiVagvolgyi07} J. van den Berg, A. A. Jarai, and B. V\'agv\"olgyi,
 The size of a pond in 2D invasion percolation,
  Electronic Comm. Probab. {\bf 12}, 411 (2007).

\bibitem{DamronSapozhnikovVagvolgyi09}
M. Damron, A. Sapozhnikov and B. V\'agv\"olgyi, Relations between invasion percolation and
critical percolation in two dimensions. Ann. Probab. {\bf 37}, 2297 (2009).


\bibitem{FehrKadauArajjoAndradeHerrmann11} 
E. Fehr, D. Kadau, N. A. M. Ara\'ujo, J. S. Andrade, and H. J. Herrmann,
Scaling relations for watersheds, Phys. Rev. E {\bf 84}, 036116 (2011).

\bibitem{Knecht}  C. L. Knecht, 

http://www.knechtmagicsquare.paulscomputing.com/



\end{thebibliography}
\end{document}